\documentclass[pdftex, 12pt, a4paper]{report}
\usepackage{newtxtext, newtxmath}
\usepackage[top=2cm, bottom=3cm, left=2cm, right=2cm]{geometry}
\usepackage[hidelinks]{hyperref}
\usepackage[pdftex]{graphicx}
\usepackage{booktabs}
\usepackage{setspace}
\usepackage{titlesec}
\usepackage{tocloft}
\usepackage{parskip}
\usepackage{amsmath}
\usepackage{amsfonts}
\usepackage[sorting=none]{biblatex}
\usepackage{subcaption}
\usepackage{wrapfig}

\title{\uppercase{DragGaussian: Enabling Drag-style Manipulation on 3D Gaussian Representation}}
\author{Sitian Shen}
\date{May 2024}

\begin{document}
\makeatletter
\begin{titlepage}
\begin{center}

\uppercase{\textbf{\Large{\@title}}}
\\[6cm]

\uppercase{\textbf{\large{Submitted by\\[0.2cm]\@author}}}

\textbf{\large{Beijing Institute of Technology}}

\vfill

The final year project work was carried out under the 3+1+1 Educational Framework at the National University of Singapore (Chongqing) Research Institute

\textbf{\@date}

\end{center}

\end{titlepage}
\makeatother

\setcounter{page}{1}
\pagenumbering{roman}

\chapter*{Abstract}
\addcontentsline{toc}{chapter}{Abstract}

User-friendly 3D object editing is a challenging task that has attracted significant attention recently. The limitations of direct 3D object editing without 2D prior knowledge have prompted increased attention towards utilizing 2D generative models for 3D editing. While existing methods like Instruct NeRF-to-NeRF offer a solution, they often lack user-friendliness, particularly due to semantic guided editing. In the realm of 3D representation, 3D Gaussian Splatting emerges as a promising approach for its efficiency and natural explicit property, facilitating precise editing tasks. Building upon these insights, we propose DragGaussian, a 3D object drag-editing framework based on 3D Gaussian Splatting, leveraging diffusion models for interactive image editing with open-vocabulary input. This framework enables users to perform drag-based editing on pre-trained 3D Gaussian object models, producing modified 2D images through multi-view consistent editing. Our contributions include the introduction of a new task, the development of DragGaussian for interactive point-based 3D editing, and comprehensive validation of its effectiveness through qualitative and quantitative experiments.

\pagebreak
\addcontentsline{toc}{chapter}{Contents}
\tableofcontents

\pagebreak
\addcontentsline{toc}{chapter}{List of Figures}
\listoffigures

\cleardoublepage
\pagenumbering{arabic}

\chapter{Introduction}

 Due to the lack of large scale 3D dataset, editing directly on 3D object or scene without 2D prior often lacks generalizability. 3D editing utilizing 2D generative models ~\cite{yoo2023plausible} has gain extensive attention recently. A famous framework of three editing was proposed in Instruct NeRF-to-NeRF~\cite{haque2023instruct}, that is to edited the multi-view projected 2D images from a 3D representation using 2D editing methods, and then use the edited 2D images as new training data to finetune the initial 3D representation. However, the 2D editing algorithm in this work called Instruct Pix2Pix~\cite{brooks2023instructpix2pix} is a semantic guided editing method, which isn't very user-friendly. Compared with semantic guidance, some dragging can better represent the user's editing goal.

To address this issue, we tried to leverage the widely adopted 2D drag editing approach known as DragGAN~\cite{pan2023draggan}, renowned for its ability to facilitate interactive point-based image manipulation. Within this framework, users initiate the editing process by selecting pairs of anchor and destination points on an image. Subsequently, the model executes semantically coherent modifications, seamlessly relocating the content from the anchor points to their corresponding destinations. Moreover, users have the option to delineate specific editable regions by drawing masks, ensuring targeted alterations while preserving the remainder of the image. However, it's worth noting that due to its reliance on GAN~\cite{GAN} within its network architecture, this method cannot deal with open-vocabulary inputs, posing a significant drawback on generalizability for its practical application in real-world scenarios. In this case, we turned to DragDiffusion, the first interactive point-based image editing method with diffusion models. Empowered by large-scale pre-trained diffusion models ~\cite{saharia2022photorealistic}, DragDiffusion achieves accurate spatial control in image editing with significantly better generalizability, while allowing open-vocabulary input data.

When it comes to 3D representation, Neural radiance field (NeRF) methods~\cite{barron2021mip, barron2022mip} have shown great power and synthesizing novel-view images. However, as current 2D diffusion models have the limited ability to localize editing regions, it is hard to generate delicate edited scenes as the unwanted regions are usually changed with diffusion models. Recent 3D Gaussian Splatting~\cite{kerbl20233d}(3D-GS) has been a groundbreaking work in the field of radiance field, which is the first to achieve a real sense of real-time rendering while enjoying high rendering quality and training speed. Besides its efficiency, we further notice its natural explicit property. 3D-GS has a great advantage for editing tasks as each 3D Gaussian exists individually. It is easy to edit 3D scenes by directly manipulating 3D Gaussians with desired constraints applied, including choosing explicit editing mask and choosing the center of one of the Gaussian points as the editing start points (handle points). Some works~\cite{chen2023gaussianeditor, fang2023gaussianeditor} have proposed effective semantic guided editing on 3D Gaussian recently.

Taking into account the above considerations, we propose a 3D object drag-editing framework based on 3D Gaussian Splatting representation, named DragGaussian. In our pipeline, users input pre-trained 3D Gaussian object models into the system through a user interface. Upon visualization, they can designate initial and final points for drag editing or specify regions for editing. These points are projected onto 2D images from various camera angles via the projection module. Subsequently, employing Multi-view Consistent Editing, we produce the modified 2D images. Prior to this, to enhance the adaptability of the pre-trained 2D editing network to diverse inputs, we fine-tune the editing network using an enhanced version of multi-view LoRA. Finally, we refine the original 3D Gaussian model using the modified 2D images to demonstrate the edited appearance of the 3D objects.

Our contributions are summarized as follows: 

1) We proposed a new task: 3D object editing through drag operation on 3D Gaussian. 

2) We present a novel 3D object editing method DragGaussian, the first to achieve interactive point-based 3D editing with diffusion models on 3D Gaussian representation. 

3) Comprehensive qualitative and experiments demonstrate the effectiveness of our DragGaussian.

\chapter{Background}
\section{Preliminaries on 3D Gaussian Splatting}
3D Gaussian splatting\cite{kerbl20233d} is a modern and effective method for 3D visualization. It uses point-based 3D Gaussians, labeled as
$G = \{g_1, g_2, \dots g_N\}$, where each $g_i$ consists of a set $\{\mu, \Sigma, c, \alpha\}$ with $i$ ranging from 1 to $N$. Here, $\mu$ represents the 3D position of the Gaussian's center, $\Sigma$ is the 3D covariance matrix, $c$ is the RGB color value, and $\alpha$ denotes the opacity. This method is notable for its streamlined Gaussian representation and its efficient, differentiable rendering technique, allowing for high-quality, real-time rendering. The rendering process of splatting is described by the equation 
\begin{equation}
    C= \sum_{i\in N} c_i\sigma_i \prod_{j=1}^{i-1}(1-\alpha_j),
\end{equation}
 where $\sigma_i = \alpha_i e^{-\frac{1}{2}(x_i)^T \Sigma^{-1}(x_i)}$ indicates the Gaussian's impact on a pixel, with $x_i$ being the pixel's distance from the $i$-th Gaussian's center.

\section{Preliminaries on Diffusion Models}
\label{DDPM&DDIM}
Denoising diffusion probabilistic models (DDPM) ~\cite{ho2020denoising} constitutes a family of latent generative models. Concerning a data distribution $q(\mathbf{z})$, DDPM approximates $q(\mathbf{z})$ as the marginal $p_\theta(\mathbf{z}_0)$ of the joint distribution between $\mathbf{Z}_0$ and a collection of latent random variables $\mathbf{Z}_{1:T}$. Specifically,

\begin{equation}
p_\theta(\mathbf{z}_0) = \int p_\theta(\mathbf{z}_0:\mathbf{z}_{1:T}) \, d\mathbf{z}_{1:T},
\end{equation}

where $p_\theta(\mathbf{z}_T)$ is a standard normal distribution and the transition kernels $p_\theta(\mathbf{z}_t | \mathbf{z}_{t-1})$ of this Markov chain are all Gaussian conditioned on $\mathbf{z}_t$. In our context, $\mathbf{Z}_0$ corresponds to image samples given by users, and $\mathbf{Z}_t$ corresponds to the latent after $t$ steps of the diffusion process. ~\cite{LDM} proposes Latent Diffusion Model (LDM), which maps data into a lower-dimensional space via a Variational Auto-Encoder (VAE) ~\cite{VAE} and models the distribution of the latent embeddings instead. Based on the framework of LDM, several powerful pretrained diffusion models have been released publicly, including the Stable Diffusion (SD) model (\url{https://huggingface.co/stabilityai}). In SD, the network responsible for modeling $p_\theta(\mathbf{z}_t | \mathbf{z}_{t-1})$ is implemented as a UNet ~\cite{unet} that comprises multiple self-attention and cross-attention modules \cite{vaswani2017attention}.

Based on DDPM method, Denoising Diffusion Implicit Models (DDIM)~\cite{song2020denoising} has also been proposed, who's sampling process is shown in Equ.~\ref{Equ:DDIM}. DDIM perfectly solves the two issues of DDPM. DDIM has a fast sampling speed, and its sampling process is deterministic because $\sigma$ can be set to 0. Therefore, no noise is introduced during the sampling process.

\begin{equation}
\label{Equ:DDIM}
x_{t-1} = \sqrt{\Bar{\alpha_{t-1}}} \left(\frac{ x_{t} - \sqrt{1 - \sqrt{\Bar{\alpha_t}}}\varepsilon_t}{\sqrt{\Bar{\alpha_t}}} \right) + \sqrt{1 - \Bar{\alpha_{t-1}} - \sigma^2\varepsilon_t} + \sigma^2 \varepsilon 
\end{equation}

The above formula can be reversed to derive the calculation formula for $x_t$ from $x_{t-1}$ as shown in Equ.~\ref{Equ:DDIM Inversion}, called DDIM Inversion. Given the initial image $x_0$, all that is needed is to use a pre-trained U-net model to predict noise, and then continuously update to obtain $x_t$.

\begin{equation}
\label{Equ:DDIM Inversion}
x_t = \frac{\sqrt{\Bar{\alpha_t}}}{\sqrt{\Bar{\alpha_{t-1}}}} (x_{t-1} - \sqrt{1 - \Bar{\alpha_{t-1}}}\varepsilon_t) + \sqrt{1 - \Bar{\alpha_t}}\varepsilon_t.
\end{equation}

\section{Preliminaries on Multi-view Diffusion Model}
\label{MVDream}

By leveraging insights from both 2D and 3D datasets, a multi-view diffusion model can attain the generalizability of 2D diffusion models while maintaining the consistency of 3D renderings. In an earlier study, MVDream~\cite{shi2023mvdream} laid the groundwork by effectively training a multi-view diffusion network. This network generates four orthogonal and coherent multi-view images based on a given text prompt and corresponding camera embedding. In our DragGaussian approach, we utilize the U-net architecture pretrained in MVDream as a foundational model.

In the training stage of MVDream, each block of the multi-view network contains a densely connected 3D attention on the four view images, which allows a strong interaction in learning the correspondence relationship between different views. To train such a network, it adopts a joint training with the rendered dataset from the Objaverse~\cite{deitke2023objaverse} and a larger scale text-to-image (t2i) dataset, LAION5B~\cite{schuhmann2022laion}, to maintain the generalizability of the fine-tuned model. Formally, given text-image dataset $\mathcal{X} = \{\mathbf{x}, y\}$ and a multi-view dataset $\mathcal{X}_{mv} = \{\mathbf{x}_{mv}, y, \mathbf{c}_{mv}\}$, where $x$ is a latent image embedding from VAE~\cite{VAE}, $y$ is a text embedding from CLIP~\cite{CLIP}, and $\mathbf{c}$ is their self-designed camera embedding, we may formulate the multi-view (MV) diffusion loss as,

\begin{equation}
\mathcal{L}_{MV}(\theta, \mathcal{X}, \mathcal{X}_{mv}) = \mathbb{E}_{\mathbf{x},y,\mathbf{c},t,\varepsilon} \left[ \left\| \varepsilon - \varepsilon_{\theta}(x^p ; y, \mathbf{c}^p, t) \right\|^2 \right]    
\end{equation}

\begin{center}
where, $(\mathbf{x}^p, \mathbf{c}^p) = 
\begin{cases} 
(\mathbf{x}, \mathbf{0}) & \text{with probability } p, \\
(\mathbf{x}_{mv}, \mathbf{c}_{mv}) & \text{with probability } 1 - p.
\end{cases}
$    
\end{center}

Here, $\mathbf{x}$ is the noisy latent image generated from a random noise $\varepsilon$ and image latent, $\varepsilon_{\theta}$ is the multi-view diffusion (MVDiffusion) model parameterized by $\theta$.

\chapter{Literature Review}

\section{2D Editing}
In the domain of 2D image editing, guidance modalities encompass visual, textual, and other forms such as point-based mouth tracking~\cite{zhan2023multimodal}. Visual guidance leverages pixel-space properties to afford precise control, allowing for the repurposing of image synthesis techniques across a range of editing tasks by adjusting visual elements like semantic maps, as seen in~\cite{zhan2022bi, zheng2022semantic}. Textual guidance, in contrast, offers a more dynamic and flexible medium for articulating visual concepts, exemplified by methods like Instruct Pix2Pix\cite{brooks2023instructpix2pix}.

Recent advancements have spotlighted point-based editing as a method for nuanced image content manipulation, illustrated by~\cite{umetani2022user, pan2023draggan, wang2022rewriting}. Specifically, DragGAN~\cite{pan2023draggan} employs latent code optimization and point tracking for effective dragging manipulation, though its utility is bounded by the intrinsic limitations of GANs. FreeDrag~\cite{ling2023freedrag} seeks to enhance DragGAN through a novel point-tracking-free approach. Furthermore, advancements in diffusion models~\cite{ho2020denoising} have led to developments such as DragDiffusion~\cite{shi2023dragdiffusion} and DragonDiffusion~\cite{mou2023dragondiffusion}, which adapt DragGAN's framework to the diffusion model context, significantly enhancing its versatility and generalization capabilities.

\section{3D Editing}

In the realm of NeRF-based 3D editing, the integration of text and image-guided methodologies marks a significant advancement. Works such as Sine~\cite{bao2023sine}, NeRF-Art~\cite{wang2023nerf}, and TextDeformer~\cite{gao2023textdeformer} employ semantic-driven and geometric manipulation techniques, while Clip-NeRF~\cite{wang2022clip} innovatively combines text and image inputs for enhanced NeRF manipulation. Building on these foundations, InstructNeRF2NeRF~\cite{haque2023instruct} introduces a novel text-guided scene editing approach using 2D diffusion models, albeit with the caveat of potential global scene impacts due to its 2D image reliance. In parallel, methods like Ed-NeRF~\cite{park2023ed} and DreamEditor~\cite{zhuang2023dreameditor} utilize static masking to define editing zones, and Watch Your Steps~\cite{mirzaei2023watch} dynamically localizes edits via relevance mapping during NeRF training, further refining the precision of NeRF-based editing.

Focusing on object-level modifications, FocalDreamer~\cite{li2023focaldreamer} offers meticulous control over object attributes, ensuring detailed and consistent alterations, while Image Sculpting~\cite{yenphraphai2024image} introduces an intuitive platform for geometric adjustments, enabling users to artistically sculpt and modify objects within 3D spaces. APAP~\cite{yoo2023plausible} offers a novel shape deformation method, enables plausibility-aware mesh deformation and preservation of fine details of the original mesh while offering an interface that alters geometry by directly displacing a handle along a certain direction. These approaches exemplify the nuanced control and artistic freedom now achievable in object editing.

Expanding the spectrum of 3D editing techniques, Gaussian-based methods, as exemplified by GaussianEditor~\cite{fang2023gaussianeditor, chen2023gaussianeditor}, utilize text directives for the refined manipulation of 3D Gaussians. This underscores the agility, precision, and controllability inherent in Gaussian representations for 3D edits. In a complementary vein, Drag3D merges the capabilities of DragGAN\cite{pan2023draggan} and Get3D~\cite{gao2022get3d} to pioneer an interactive 3D dragging edit demonstration. This mesh-based manipulation technique empowers users with direct, intuitive interactions, leveraging advanced generative and reconstructive technologies to facilitate object transformations, thereby enriching the toolkit available for 3D object editing. However, Drag3D's mesh-based nature can result in less smooth and realistic 3D representations, and its reliance on DragGAN restricts it to predefined semantics, limiting open-ended drag editing.

\section{Multi-view Diffusion Model}
Traditional 2D diffusion models\cite{rombach2022high, saharia2022photorealistic} are tailored for single-view image generation, lacking capabilities for 3D viewpoint adjustments. In response, recent innovations such as SweetDreamer~\cite{li2023sweetdreamer}, Wonder3D~\cite{long2023wonder3d}, Zero123++~\cite{shi2023zero123++},  Viewset Diffusion~\cite{szymanowicz2023viewset}, SyncDreamer~\cite{liu2023syncdreamer}, Mvdream~\cite{shi2023mvdream}, and Imagedream~\cite{wang2023imagedream} have emerged, refining multi-view diffusion models with 3D data to integrate camera poses, thereby facilitating multi-view image generation of a singular object from text prompts or single images. Concurrently, 3D editing techniques like Efficient-NeRF2NeRF~\cite{song2023efficient} have leveraged multi-view enhanced diffusion models to heighten editing precision.
\chapter{Methodology}

In this section, we formally present the proposed DragGaussian pipeline and the whole system. The whole pipeline is shown in Fig.~\ref{fig:example}. Users upload pre-trained 3D Gaussian object models into the system via a UI interface, whereupon visualization, they can select starting and ending points for dragging editing or editable regions (Sec.~\ref{UI}). Through the projection module, we obtain the coordinates of the selected points on 2D images from different camera perspectives. Subsequently, via Multi-view Consistent Editing (Sec.~\ref{Edit}), we generate the edited 2D images. Prior to this, to better adapt the pre-trained 2D editing network to the open-vocabulary input, we fine-tune the editing network using an improved version of multi-view LoRA (Sec.~\ref{LoRA}). Finally, we further train the original 3D Gaussian model using the edited 2D images to present the edited effect of the 3D objects.

\begin{figure}[h]
    \centering
  \includegraphics[width=\textwidth]{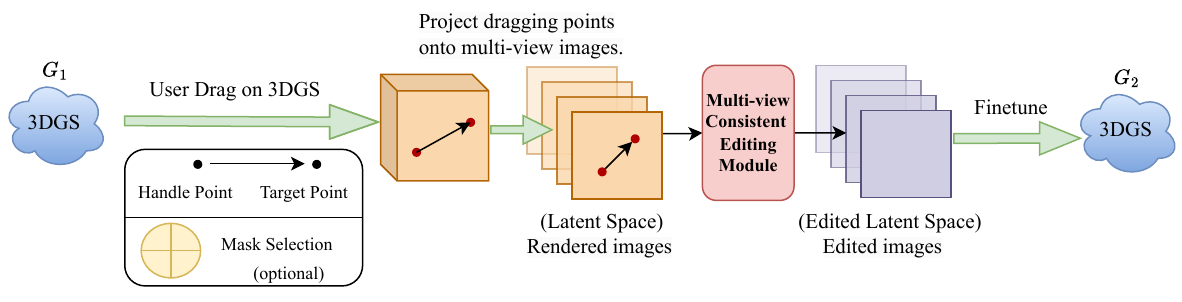}
  \caption{Pipeline of DragGaussian.
  }
  \label{fig:example}
\end{figure}

\section{Interactive 3D Point-based Manipulation}
\label{UI}

\begin{figure}[h]
    \centering
    \begin{subfigure}{0.47\linewidth}
        \centering
        \includegraphics[width=\linewidth]{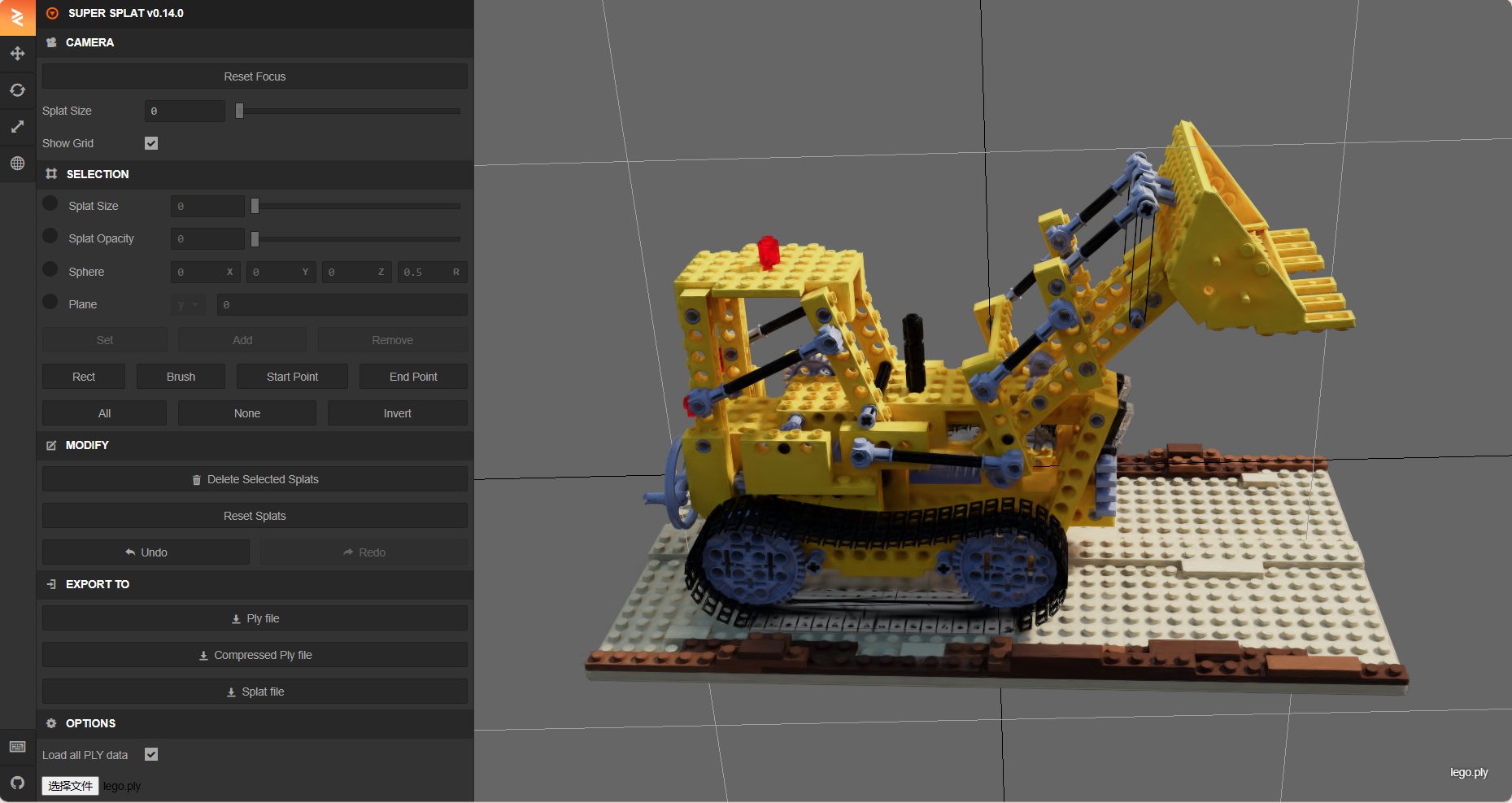}
        \caption{Splat size configured to 0.}
        \label{subfig:left}
    \end{subfigure}
    \hfill
    \begin{subfigure}{0.47\linewidth}
        \centering
        \includegraphics[width=\linewidth]{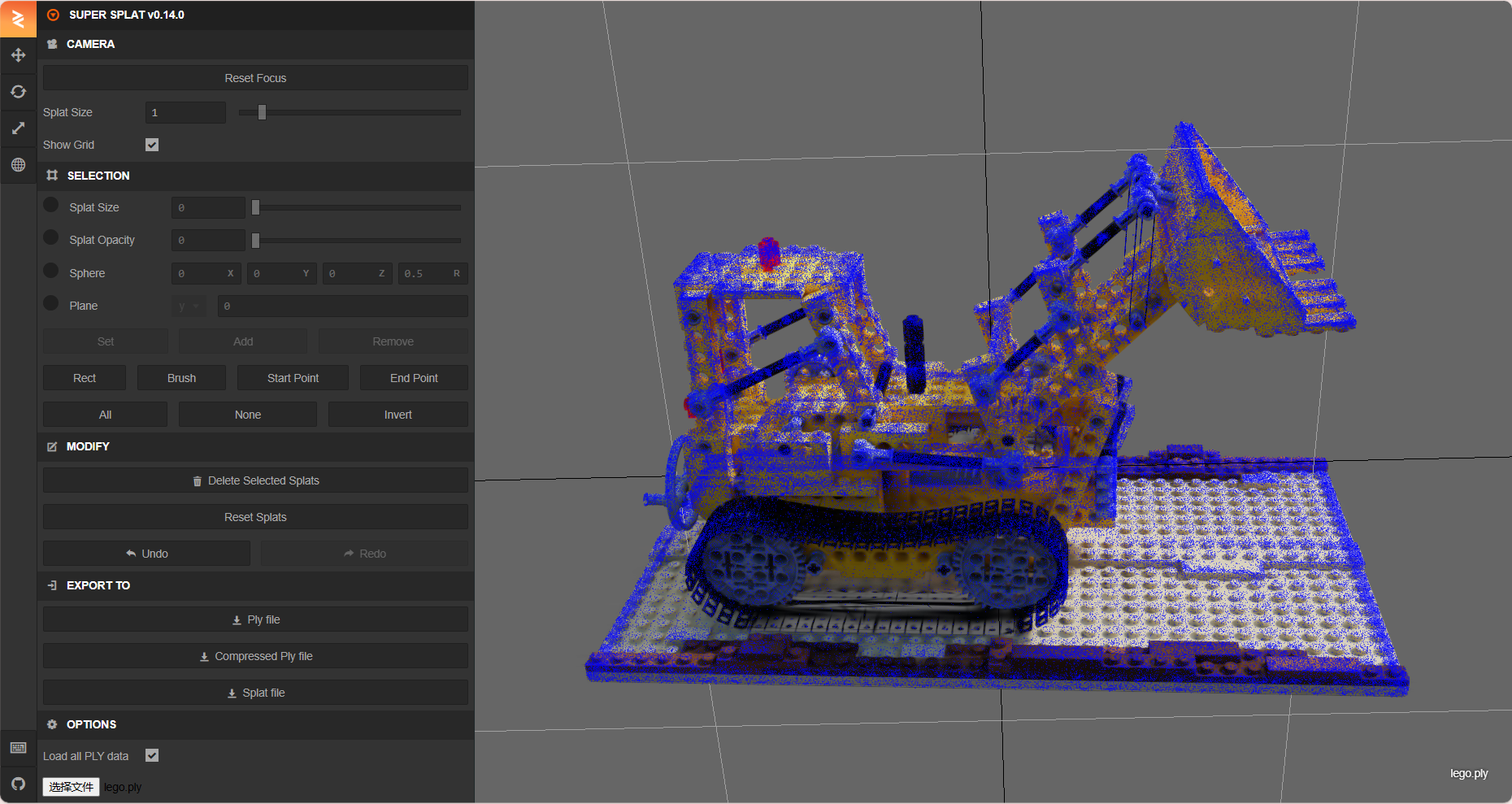}
        \caption{Spat size configured to 1.}
        \label{subfig:right}
    \end{subfigure}
    \caption{Overview of our UI.}
    \label{fig:UI}
\end{figure}

The user interface (UI) overview is illustrated in Fig.~\ref{fig:UI}. Leveraging an initial UI repository, we have developed our custom supersplat tool. Several buttons, including 'start points,' 'end points,' and 'brush,' have been integrated into this interface. Users have the flexibility to select a precise $n$ Gaussian points as start points for dragging ($\mathbf{S}={S_1, S_2, ..., S_n}$), followed by the selection of corresponding end points anywhere within the 3D space ($\mathbf{E}={E_1, E_2, ..., E_n}$). Subsequently, the coordinates of these points, along with four randomly chosen camera poses $\mathbf{C}=\{C_1, C_2, C_3, C_4\}$, are passed to a projection module $P$. The projection module computes the coordinates of these points on 2D splatted images of the 3D Gaussian from specific poses, denoted as $\mathbf{s}=\{s_1, s_2, ..., s_n\}$ and $\mathbf{e}=\{e_1, e_2, ..., e_n\}$, using the following functions:
\begin{equation}
    \mathbf{s} = P(\mathbf{S}, \mathbf{C}), \mathbf{e} = P(\mathbf{E}, \mathbf{C}).
\end{equation}

Different from other 3D object representation methods, the 3D Gaussian is explicit, allowing users to directly specify editable Gaussians in 3D space. In our UI, users can click the 'brush' button to apply a mask to Gaussian points eligible for editing, as illustrated in Fig. 3. This set of Gaussian points, $G_{mask}$, which is subject to updates, will be immediately fed into the model. During the subsequent fine-tuning process of pre-trained Gaussians, $G_{mask}$ serves as a control mechanism, ensuring that parameters of Gaussians not belonging to $G_{mask}$ are not updated.

\begin{figure}[h]
    \centering
    \begin{subfigure}{0.47\linewidth}
        \centering
        \includegraphics[width=\linewidth]{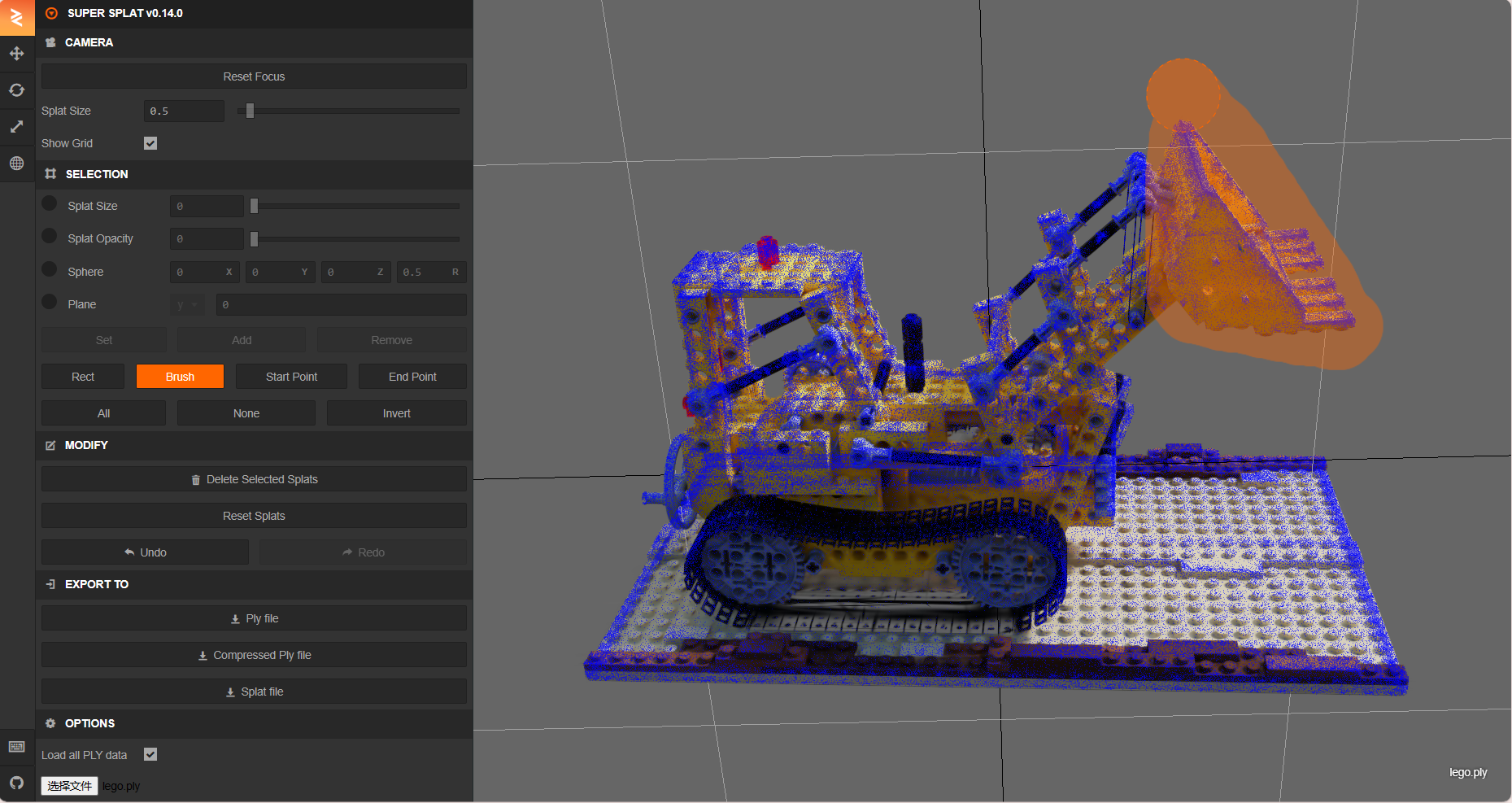}
        \caption{Splat size configured to 0.}
        \label{subfig:left}
    \end{subfigure}
    \hfill
    \begin{subfigure}{0.47\linewidth}
        \centering
        \includegraphics[width=\linewidth]{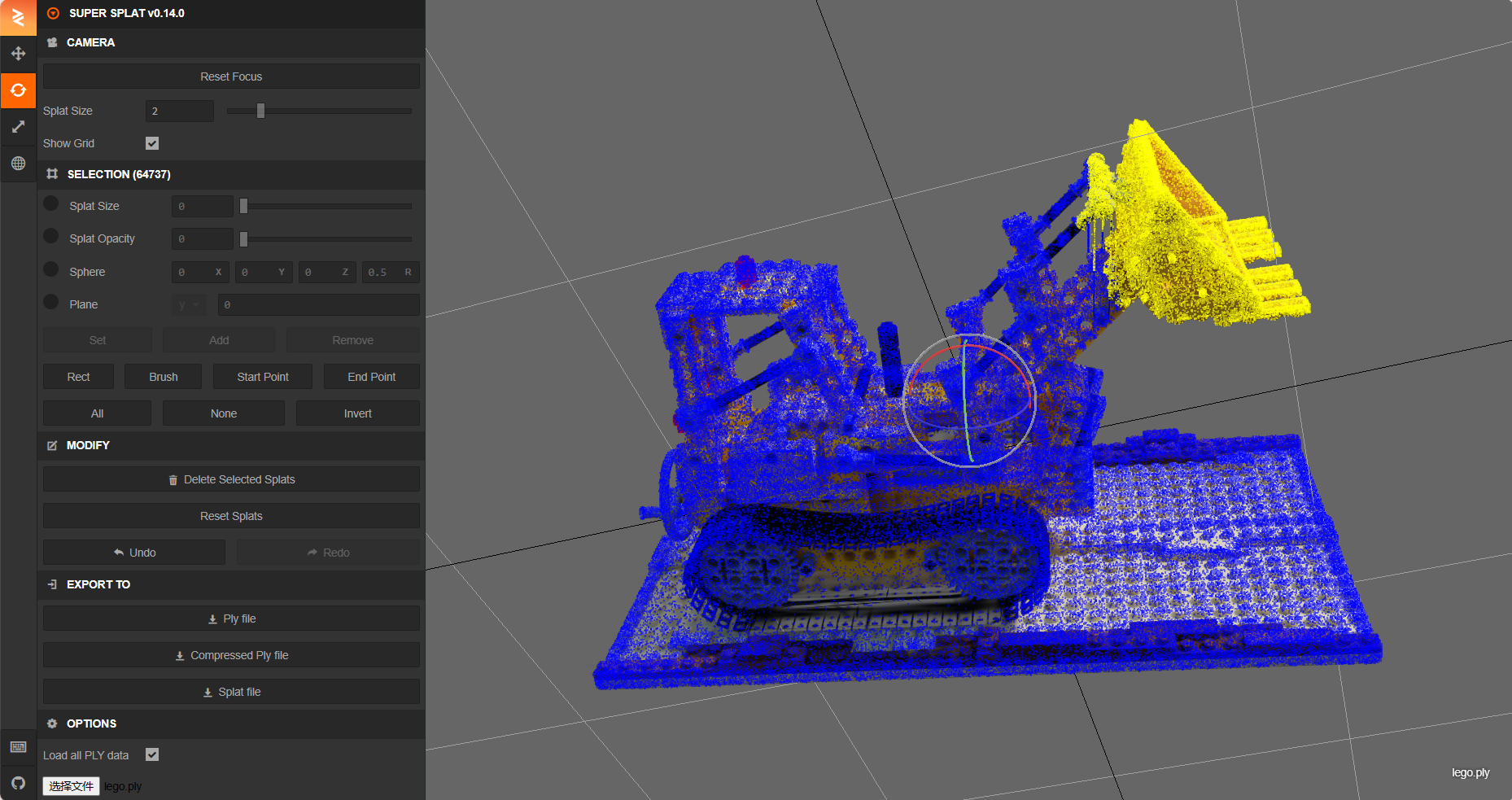}
        \caption{Spat size configured to 1.}
        \label{subfig:right}
    \end{subfigure}
    \caption{Drawing mask using the brush.}
    \label{fig:mask}
\end{figure}

\section{Multi-view Identity Preserving Fine-tuning}
\label{LoRA}
Before editing on multi-view 2d images, we conduct an extra multi-view identity-preserving fine-tuning on the pre-trained multi-view diffusion U-Net. This stage aims to ensure that the diffusion denoising U-Net encodes the features of multi-view images more accurately (than in the absence of this procedure), thus facilitating the consistency of the identity of the multi-view images throughout the editing process. Following the fine-tuning method in DragDiffusion, our fine-tuning process is also implemented with LoRA~\cite{hu2021lora}, whose objective function is
\begin{equation}
\mathcal{L}_{\text{ft}}(z, \Delta\theta) = \mathbb{E}_{\epsilon,t}\left[\left\|\epsilon - \epsilon_{\theta + \Delta\theta}(\alpha_t z + \sigma_t \epsilon)\right\|_2^2\right],    
\end{equation}
where \(\theta\) and \(\Delta\theta\) represent the U-Net and LoRA parameters respectively, \( z \) is the real image, \( \epsilon \sim \mathcal{N}(0, \mathbf{I}) \) is the randomly sampled noise map, \( \epsilon_{\theta + \Delta\theta}(\cdot) \) is the noise map predicted by the LoRA-integrated U-Net, and \( \alpha_t \) and \( \sigma_t \) are parameters of the diffusion noise schedule at diffusion step \( t \). The fine-tuning objective is optimized via gradient descent on \( \Delta\theta \). 

During fine-tuning, images of $n$ views from the same 3D gaussian are combined into a single batch, noted as $\mathbf{z}$, sharing a same sampled timestep $t$ and independently sampled noise maps $\epsilon_i$.
For each batch, our loss function can be specified as:
\begin{equation}
\mathcal{L}_{\text{batch}}(\mathbf{z}, \Delta\theta) = \frac{1}{n}\sum_{i=1}^n\left[\left\|\epsilon_i - \epsilon_{\theta + \Delta\theta}(\alpha_t \mathbf{z}_i + \sigma_t \epsilon_i)\right\|_2^2\right].    
\end{equation}
Following the setting of MVDream, we set $n=4$ during fine-tuning.

\section{Multi-view Consistent Editing}
\label{Edit}

\begin{figure}[h]
    \centering
  \includegraphics[width=\textwidth]{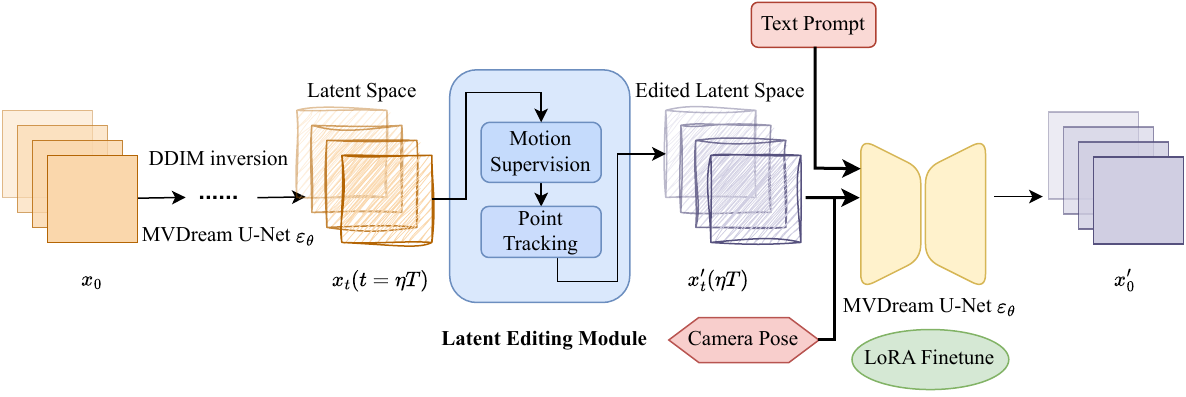}
  \caption{Stages for Multi-view Consistent Editing.}
  \label{fig:example}
\end{figure}

After multi-view identity preserving fune-tuning, we add noise to the projected 2d images using DDIM inversion (Sec.~\ref{DDPM&DDIM}), and then implement motion supervision and point tracking methods to optimize the diffusion latent according to the user directions to achieve editing on 2D images. 

\subsection{Motion Supervision}

We denote the $n$ 2d handle points (start points) at the $k$-th motion supervision iteration as $\{h_i^k = (x_i^k, y_i^k) : i = 1, \ldots, n\}$ and their corresponding target points as $\{e_i^k = (x_i^k, y_i^k) : i = 1, \ldots, n\}$. Each of the input four images is denoted as $z_0$; the $t$-th step latent (the result of $t$-th DDIM inversion) is denoted as $z_t$. 
We pass $z_t$ through the U-Net architecture extracted from the pre-trained MVDream model. Let $F(z_t)$ represent the output feature maps of the U-Net used for motion supervision, and $F_{h_i^k}(z_t)$ denote the feature vector at pixel location $h_i^k$.
Also, we denote the square patch centered around $h_i^k$ as $\Omega(h_i^k; r_1) = \{(x, y) : |x-x_i^k| \leq r_1, |y-y_i^k| \leq r_1\}$. Then, the motion supervision loss at the $k$-th iteration is defined as:

\begin{equation}
\label{equ:motion_supervision}
\begin{aligned}
L_{ms}(z_t^k) &= \sum_{i=1}^{n} \sum_q \Omega \left( \left\| F_{q + d_i}(\hat{z}_t^k) - sg(F_q(\hat{z}_t^k)) \right\|_1 \right. \\
&\quad + \lambda \left\| \left( \hat{z}_t^k - sg(\hat{z}_{t-1}^0) \right) \odot \left(1 - M\right) \right\|_1 ,
\end{aligned}
\end{equation}

where $\hat{z}_t^k$ is the $t$-th step latent after the $k$-th update, $\text{sg}(\cdot)$ is the stop gradient operator (i.e., the gradient will not be back-propagated for the term $\text{sg}(F_q(\hat{z}_t^k)))$, $d_i = \left(\frac{{g_i - \hat{h}_k^i}}{{\|g_i - \hat{h}_k^i\|_2}}\right)$, $\|g_i - \hat{h}_k^i\|_2$ is the normalized vector pointing from $\hat{h}_k^i$ to $g_i$, $M$ is the binary mask specified by the user, $F_{q+d_i}(\hat{z}_t^k)$ is obtained via bilinear interpolation as the elements of $q + d_i$ may not be integers. In each iteration, $\hat{z}_t^k$ is updated by taking one gradient descent step to minimize $L_{ms}$:

\begin{equation}
\hat{z}_t^{k+1} = \hat{z}_t^k - \eta \frac{{\partial L_{ms}(\hat{z}_t^k)}}{{\partial \hat{z}_t^k}},    
\end{equation}

where $\eta$ is the learning rate for latent optimization.

\subsection{Point Tracking}

Since the motion supervision updates $\hat{z}_t^k$ on the four 2d images, the positions of the handle points on them may also change. Therefore, we need to perform point tracking on each of the four images to update the handle points after each motion supervision step. To achieve this goal, we use Multi-view U-Net feature maps $F(\hat{z}_t^{k+1})$ and $F(\hat{z}_t)$ to track the new start points. Specifically, we update each of the start points $h_i^k$ with a nearest neighbor search within the square patch $\Omega(h_i^k;r2) = \{(x,y) : |x - x_i^k| \leq r2, |y - y_i^k| \leq r2\}$ as follows:

\begin{equation}
\label{equ:point_tracking}
h_i^{k+1} = \text{arg min}_{q \in \Omega(h_i^k;r2)} \|F_q(\hat{z}_t^{k+1}) - F_{h_i^0}(\hat{z}_t)\|_1.    
\end{equation}

\subsection{Multi-view Consistent Denoising}

After we get editted 2d images from several viewpoints (here we set it as 4), we come to the stage to denoise them. Implementing the U-net pre-trained by MVDream (Sec.~\ref{MVDream}) and finetuned by LoRA (Sec.~\ref{LoRA}) to predict noise, following the equation of DDIM Sampling (Equ.~\ref{Equ:DDIM}), the four edited latent 2d images can be denoised step by step. Same as the Sampling process in DragDiffusion which uses U-net from Stable Diffusion~\cite{rombach2022high}, we replace the key and value vectors generated from $\hat{z}_t$ with the ones generated from  $z_t$. With this simple replacement technique, the query vectors generated from $\hat{z}_t$ will be directed to query the correlated contents and texture of $z_t$.

\chapter{Experiments}

\section{Implementation Details}

In our experimental setup, we configured the number of steps for DDIM Inversion and Sampling to be 50, and set the guidance scale $\eta$ to 1.0. Specifically, during the DDIM Sampling process, we shared the Key-Value (KV) pairs from layer 8 out of 10, and from step 0. When editing real images, we do not apply classifier-free guidance (CFG) ~\cite{ho2022classifier} in both DDIM Inversion and DDIM Denoising process. For the LoRA fine-tuning process, we employed a learning rate of 0.0005 for 300 LoRA steps. During the Multi-view Consistent Editing Module, we use the Adam optimizer with a learning rate of 0.01 to optimize the latent. The maximum optimization step is set to be 80. The hyperparameter $r_1$ in Eqn.~\ref{equ:motion_supervision} and $r_2$ in Equ.~\ref{equ:point_tracking} are tuned to be 1 and 3, respectively. $\lambda$ in Equ.~\ref{equ:motion_supervision} is set to 0.1 by default.
Our experiments were conducted utilizing 2 NVIDIA RTX 3090 GPUs.

\section{Editing Results}
\begin{figure}[h]
    \centering
    \begin{subfigure}{0.24\linewidth}
        \centering
        \includegraphics[width=\linewidth]{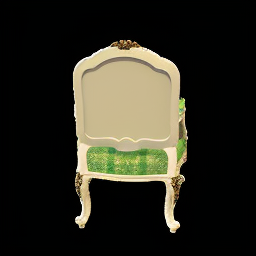}
    \end{subfigure}
    \hfill
    \begin{subfigure}{0.24\linewidth}
        \centering
        \includegraphics[width=\linewidth]{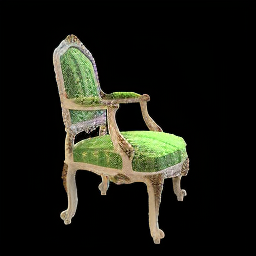}
    \end{subfigure}
    \hfill
    \begin{subfigure}{0.24\linewidth}
        \centering
        \includegraphics[width=\linewidth]{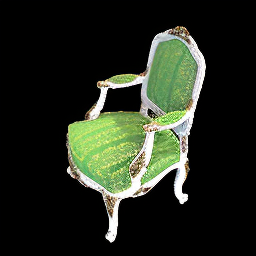}
    \end{subfigure}
    \hfill
    \centering
    \begin{subfigure}{0.24\linewidth}
        \centering
        \includegraphics[width=\linewidth]{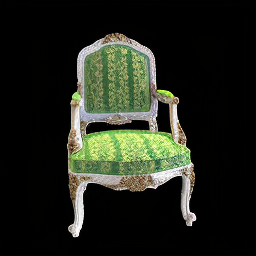}
    \end{subfigure}
    \newline
    \begin{subfigure}{0.24\linewidth}
        \centering
        \includegraphics[width=\linewidth]{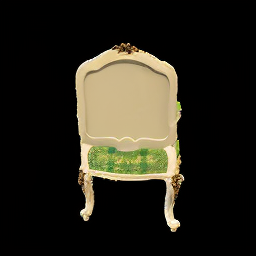}
    \end{subfigure}
    \hfill
    \begin{subfigure}{0.24\linewidth}
        \centering
        \includegraphics[width=\linewidth]{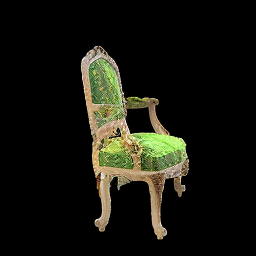}
    \end{subfigure}
    \hfill
    \centering
    \begin{subfigure}{0.24\linewidth}
        \centering
        \includegraphics[width=\linewidth]{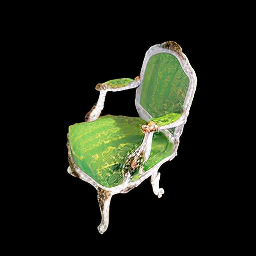}
    \end{subfigure}
    \hfill
    \begin{subfigure}{0.24\linewidth}
        \centering
        \includegraphics[width=\linewidth]{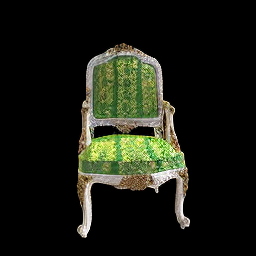}
    \end{subfigure}
    \caption{Multi-view consistent editing on a chair.}
    \label{fig:chair_2d}    
    
\end{figure}

Fig.~\ref{fig:lego_2d&3d} shows the editing results of a pretrained lego 3D Gaussian. Images on the left column are rendered from initial well-trained Gaussian, images on the middle column are 2d editing results of the model with start points (red) and end points (blue), and images on the right column are rendered results of the editted 3D Gaussian after finetuning 5000 iterations. In the editing results of both 2D and 3D, we can clearly see the shrinkage deformation caused by editing the LEGO tractor. However, constrained by the limitations of the 2D drag editing model, the resolution of the 2D editing results is lower, further leading to a blurred visual effect of the edited 3D Gaussian model.

We also compared our method with Drag3D, a 3d mesh editing methods based on dragging operation. Because of the usage of diffusion model, our method is much power than Drag3D, which extends the idea of DragGAN into GET3D. However, the representation of 3D Gaussian provides a much clearer rendering result and also provide users the possibility to choose editing areas explicitly.

\section{Ablation Study}

\begin{wrapfigure}{r}{0.3\textwidth}
\centering
\includegraphics[width=0.28\textwidth]{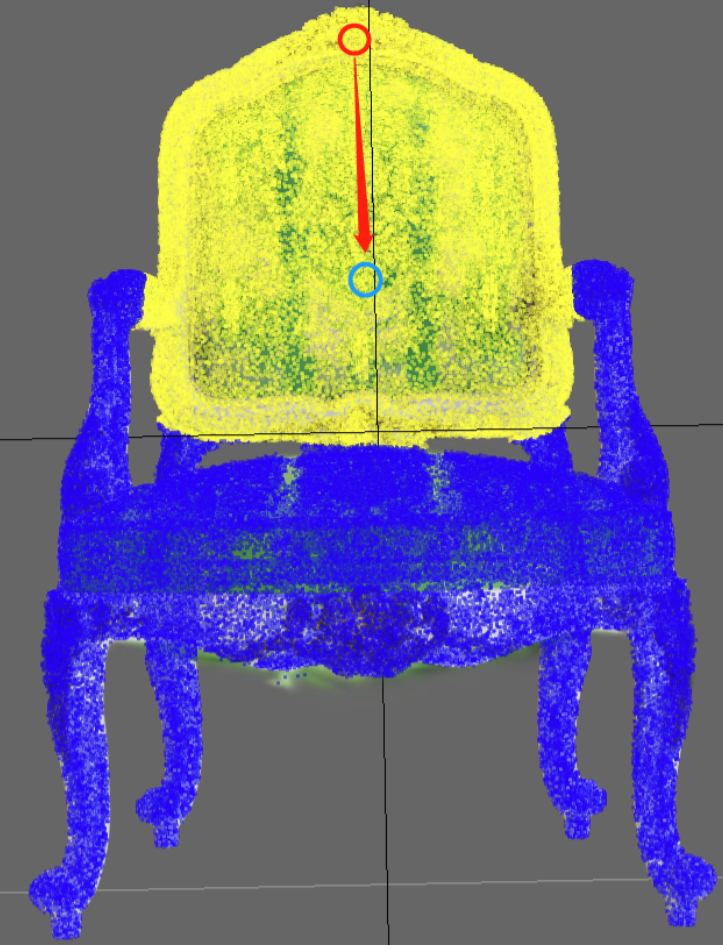}
\caption{Editing and masking on a Gaussian chair.}
\label{fig:mask_chair}
\end{wrapfigure}

To showcase the effectiveness of our editing approach for 2D images using the pre-trained MVDream network, we conducted several preliminary experiments. These experiments involved editing both multi-view 2D images derived from 3D Gaussian datasets' projections and multi-view images generated by MVDream itself. 

In Fig.\ref{fig:chair_2d}, an exemplary editing outcome is presented. The chair data originates from the NeRF Synthetic dataset and was utilized for pre-training a 3D Gaussian model. 
Fig.\ref{fig:mask_chair} demonstrates the user's editing process and mask selection, depicting the utilization of brush tools for mask creation. The editing outcome on an open vocabulary dataset reveals conspicuous traces of editing on the chair's back, along with evident distortions in unmasked areas such as the chair legs. Additionally, the overall size of the chair appears diminished post-editing. The constraints imposed by the pre-trained MVDream U-net's generalizability become apparent when confronted with unseen data, highlighting limitations in our 2D consistent editing pipeline.

\begin{figure}[h]
    \centering
    \begin{subfigure}{0.3\linewidth}
        \centering
        \includegraphics[width=\linewidth]{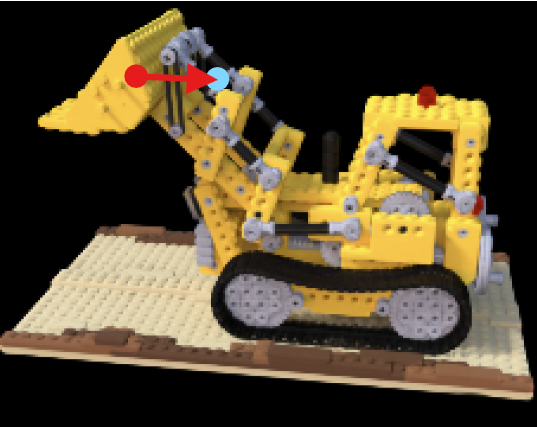}
    \end{subfigure}
    \hfill
    \begin{subfigure}{0.3\linewidth}
        \centering
        \includegraphics[width=\linewidth]{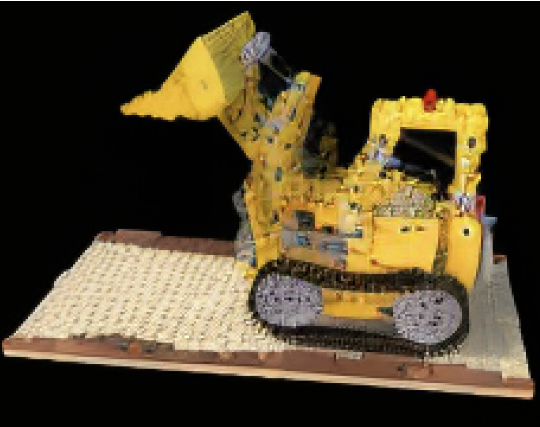}
    \end{subfigure}
    \hfill
    \begin{subfigure}{0.3\linewidth}
        \centering
        \includegraphics[width=\linewidth]{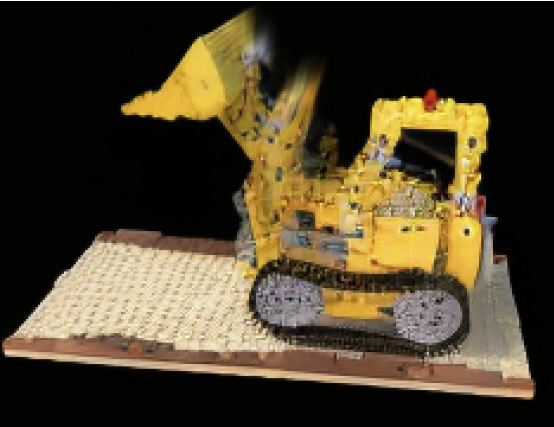}
    \end{subfigure}
    \newline
    \centering
    \begin{subfigure}{0.3\linewidth}
        \centering
        \includegraphics[width=\linewidth]{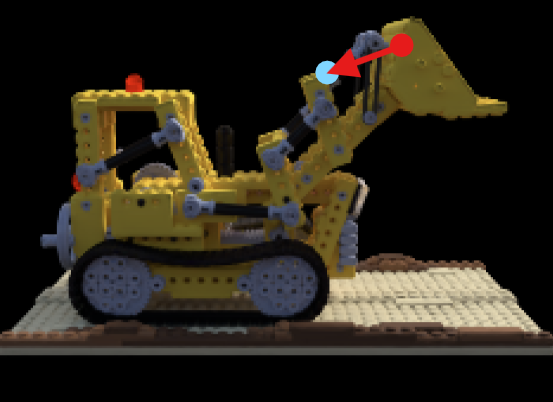}
    \end{subfigure}
    \hfill
    \begin{subfigure}{0.3\linewidth}
        \centering
        \includegraphics[width=\linewidth]{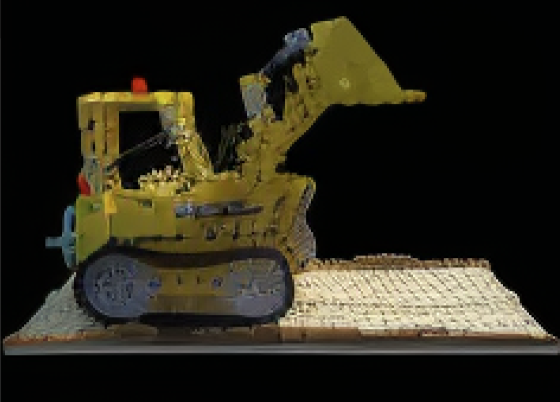}
    \end{subfigure}
    \hfill
    \begin{subfigure}{0.3\linewidth}
        \centering
        \includegraphics[width=\linewidth]{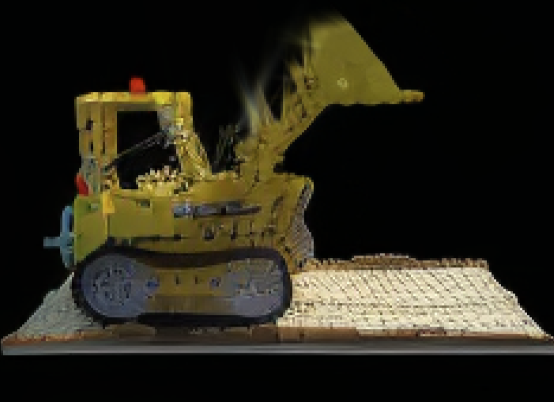}
    \end{subfigure}
    \newline
    \centering
    \begin{subfigure}{0.3\linewidth}
        \centering
        \includegraphics[width=\linewidth]{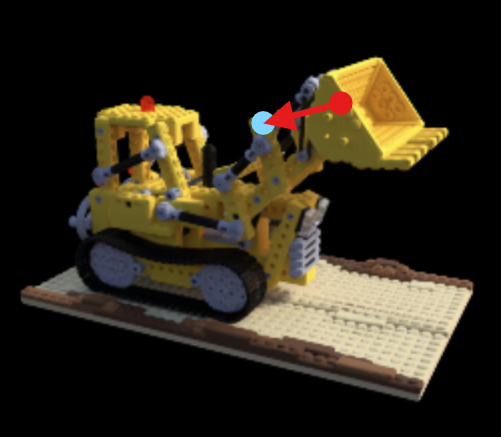}
    \end{subfigure}
    \hfill
    \begin{subfigure}{0.3\linewidth}
        \centering
        \includegraphics[width=\linewidth]{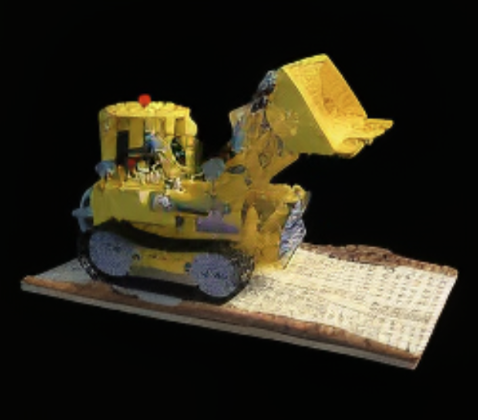}
    \end{subfigure}
    \hfill
    \begin{subfigure}{0.3\linewidth}
        \centering
        \includegraphics[width=\linewidth]{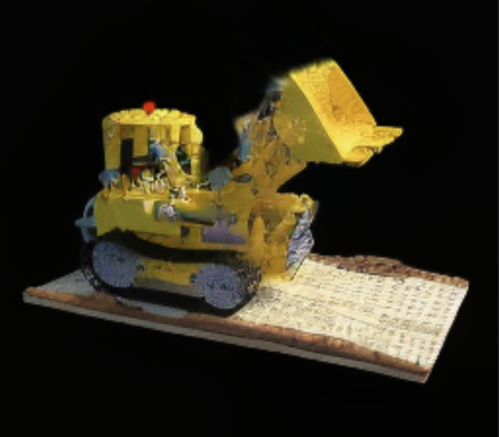}
    \end{subfigure}
    \newline
    \centering
    \begin{subfigure}{0.3\linewidth}
        \centering
        \includegraphics[width=\linewidth]{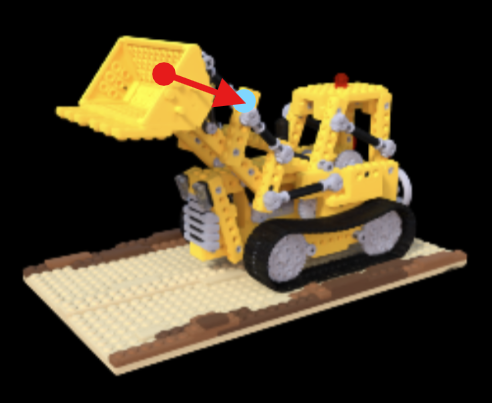}
    \end{subfigure}
    \hfill
    \begin{subfigure}{0.3\linewidth}
        \centering
        \includegraphics[width=\linewidth]{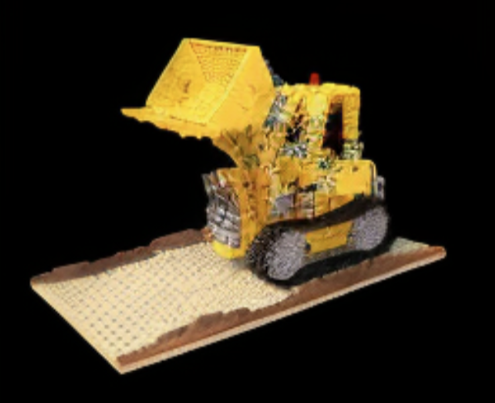}
    \end{subfigure}
    \hfill
    \begin{subfigure}{0.3\linewidth}
        \centering
        \includegraphics[width=\linewidth]{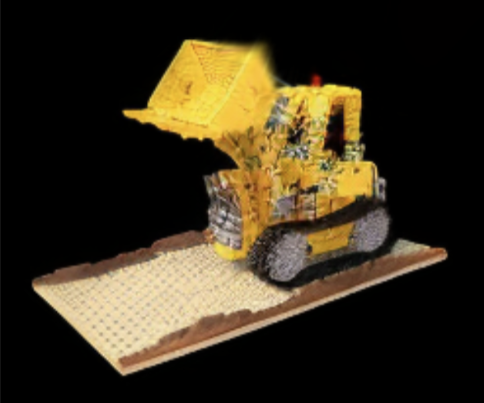}
    \end{subfigure}
    \caption{Drawing mask using the brush.}
    \label{fig:lego_2d&3d}    
    
\end{figure}

\begin{figure}[h]
    \centering
  \includegraphics[width=0.8\textwidth]{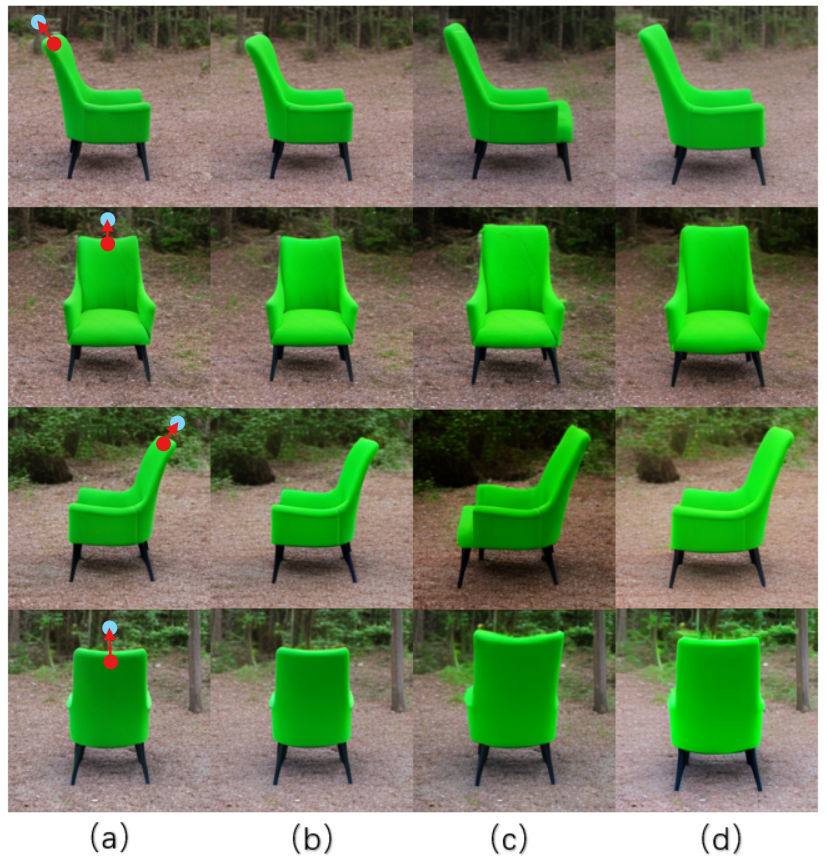}
  \caption{Editing results on multi-view images generated by MVDream. (a) column shows the initial images, (b) colunm shows the images after DDIM Inversion and DDIM Sampling process without editing, (c) column shows the images edited by our multi-view image consistent editing pipeline without LoRA finetune, and (d) column shows the editing results with LoRA finetune.}
  \label{fig:ablation}
\end{figure}

Nonetheless, our reliable 2D editing pipeline adeptly handles data generated by the U-net. When provided with a text prompt like "A chair," MVDream generates chair images from four different viewpoints. Subsequently, we feed these four chairs into DDIM Inversion, the 2D image editing process, and the multi-view denoising stage within DragGaussian's pipeline, as illustrated in Fig.\ref{Edit}. To enhance the editing process, we include four pairs of manually selected start and end points. The resulting purely 2D edited images are depicted in Fig.\ref{fig:ablation}.

To validate the effectiveness of LoRA finetuning, we contract the editing result on multi-view 2d images with and without LoRA finetune. In Fig.~\ref{fig:ablation}, (c) column shows the images edited by our multi-view image consistent editing pipeline without LoRA finetune, and (d) column shows the editing results with LoRA finetune. It's evident that editing outcomes lacking LoRA fine-tuning fail to maintain the original background color and inadvertently alter parts of the chair that weren't intended to be edited. Conversely, employing LoRA fine-tuning on the U-net of MVDream yields significantly superior results, preserving undisturbed regions more effectively.

\chapter{Discussion}

\textbf{Miss Point Tracking Issue}
The point tracking method used in our DragGaussian bases on a traditional framework for point based image editing. However, there may exist several issues in this tracking framework, which is discussed in a recent work called FreeDrag~\cite{ling2023freedrag}.

One issue is \textbf{miss tracking}, where the process of dragging points faces difficulty in accurately following the desired handle points. This problem is especially prevalent in highly curved areas with a significant perceptual path length, as observed within latent space~\cite{karras2019style}. In such instances, the optimized image undergoes significant alterations, causing handle points in subsequent iterations to be placed outside the intended search area. Furthermore, in certain cases, miss tracking can result in handle points disappearing. It's noteworthy that during miss tracking, the cumulative error in the motion supervision step progressively increases as iterations continue, due to the misalignment of tracked features. Another issue is \textbf{ambiguous tracking}, where tracked points are positioned within regions resembling the handle points. This dilemma occurs when parts of the image share features similar to the intended handle points, leading to uncertainty in the tracking process. This problem presents a significant challenge as it can mislead the motion supervision process in subsequent iterations, resulting in inaccurate or misleading guidance.

\chapter{Conclusion}

In this work, we introduced DragGaussian, an innovative 3D object drag-editing framework built upon the principles of 3D Gaussian Splatting and leveraging diffusion models for interactive image editing with open-vocabulary input. This framework empowers users to engage in drag-based editing on pre-trained 3D Gaussian object models, resulting in the production of modified 2D images through multi-view consistent editing. Our contributions encompass the inception of a novel editing task, the development of DragGaussian for interactive point-based 3D editing, and the thorough validation of its efficacy through both qualitative and quantitative experiments.

While our experiments have demonstrated the effectiveness of our 3D editing method, several limitations remain evident. Notably, the utilization of diffusion models impedes real-time editing capabilities. Moreover, constraints associated with the pre-trained MVDream network may hinder the attainment of elegant editing results across certain datasets.

Moving forward, our focus will center on the development of real-time editing frameworks directly tailored to 3D Gaussian representations. By addressing these drawbacks and refining our approach, we aim to further enhance the utility and versatility of DragGaussian, thereby advancing the landscape of 3D object editing technologies.

\printbibliography

\end{document}